\documentclass[twocolumn]{aastex63}
\usepackage{amssymb}
\usepackage{natbib}
\usepackage{hyperref}
\usepackage{color}
\usepackage{pdfcomment}

\newcommand\T{\rule{0pt}{2.6ex}}       % Top strut
\newcommand\B{\rule[-2.0ex]{0pt}{0pt}} % Bottom strut

\submitjournal{AJ}
\received{2019 November 13}
\accepted{2020 February 19}

\shorttitle{Potential Themis Family Contribution to the JFC Population}
\shortauthors{Hsieh et al.}
\begin{document}

%\begin{frontmatter}
\title{Potential Themis Family Asteroid Contribution to the Jupiter-Family Comet Population}

\correspondingauthor{Henry H.\ Hsieh}
\email{hhsieh@psi.edu}

\author[0000-0001-7225-9271]{Henry H.\ Hsieh}
\affil{Planetary Science Institute, 1700 East Fort Lowell Rd., Suite 106, Tucson, AZ 85719, USA}
\affil{Institute of Astronomy and Astrophysics, Academia Sinica, P.O.\ Box 23-141, Taipei 10617, Taiwan}

\author[0000-0001-6349-6881]{Bojan Novakovi\'c}
\affiliation{Department of Astronomy, Faculty of Mathematics, University of Belgrade, Studentski trg 16, 11000 Belgrade, Serbia}

\author[0000-0002-0906-1761]{Kevin J.\ Walsh}
\affiliation{Southwest Research Institute, 1050 Walnut St., Suite 400, Boulder, Colorado 80302, USA}

\author[0000-0002-5821-4066]{Norbert Sch{\"o}rghofer}
\affiliation{Planetary Science Institute, 1700 East Fort Lowell Rd., Suite 106, Tucson, AZ 85719, USA}

%% AASTeX 6.1 has the new \collaboration and \nocollaboration commands to
%% provide the collaboration status of a group of authors. These commands 
%% can be used either before or after the list of corresponding authors. The
%% argument for \collaboration is the collaboration identifier. Authors are
%% encouraged to surround collaboration identifiers with ()s. The 
%% \nocollaboration command takes no argument and exists to indicate that
%% the nearby authors are not part of surrounding collaborations.

%% Mark off the abstract in the ``abstract'' environment. 
\begin{abstract} % limit 250 words
Recent dynamical analyses suggest that some Jupiter family comets (JFCs) may originate in the main asteroid belt instead of the outer solar system. This possibility is particularly interesting given evidence that icy main-belt objects are known to be present in the Themis asteroid family.  We report results from dynamical analyses specifically investigating the possibility that icy Themis family members could contribute to the observed population of JFCs.  Numerical integrations show that such dynamical evolution is indeed possible via a combination of eccentricity excitation apparently driven by the nearby 2:1 mean-motion resonance with Jupiter, gravitational interactions with planets other than Jupiter, and the Yarkovsky effect.  
We estimate that, at any given time, there may be tens of objects from the Themis family on JFC-like orbits with the potential to mimic active JFCs from the outer solar system, although not all, or even any, may necessarily be observably active.
%Purely gravitational integrations indicate that $\sim$20 objects \addedtext{1~km in diameter or larger} from the Themis family could be on JFC-like orbits at any given time, where non-gravitational effects and other perturbations (e.g., collisions) could increase this number by a factor of two or more, although not all of these objects may necessarily be observably active.
We find that dynamically evolved Themis family objects on JFC-like orbits have semimajor axes between 3.15~au and 3.40~au for the vast majority of their time on such orbits, consistent with the strong role that the 2:1 mean-motion resonance with Jupiter likely plays in their dynamical evolution.  We conclude that a contribution from the Themis family to the active JFC population is plausible, although further work is needed to better characterize this contribution.
\end{abstract}

%% Keywords should appear after the \end{abstract} command. 
%% See the online documentation for the full list of available subject
%% keywords and the rules for their use.

%\begin{keyword}
%Asteroids; Comets, nucleus; Comets, dust
%\end{keyword}

%\end{frontmatter}

%% From the front matter, we move on to the body of the paper.
%% Sections are demarcated by \section and \subsection, respectively.
%% Observe the use of the LaTeX \label
%% command after the \subsection to give a symbolic KEY to the
%% subsection for cross-referencing in a \ref command.
%% You can use LaTeX's \ref and \label commands to keep track of
%% cross-references to sections, equations, tables, and figures.
%% That way, if you change the order of any elements, LaTeX will
%% automatically renumber them.

%% We recommend that authors also use the natbib \citep
%% and \citet commands to identify citations.  The citations are
%% tied to the reference list via symbolic KEYs. The KEY corresponds
%% to the KEY in the \bibitem in the reference list below. 

\section{INTRODUCTION}\label{section:introduction}
\subsection{Background}\label{section:background}

The Themis asteroid family has come to be of particular interest in solar system science in recent years. At least three main-belt comets \citep[MBCs, which exhibit comet-like activity indicative of sublimating ice, yet have asteroid-like orbits;][]{hsieh2006_mbcs}, namely 133P, 176P, and 288P, are members of the family \citep{hsieh2018_activeastfamilies}. A fourth MBC, 238P, has also been proposed to be a past member of the family \citep{haghighipour2009_mbcorigins}. Water ice frost has also been reported on large Themis family members, (24) Themis and (90) Antiope \citep{rivkin2010_themis,campins2010_themis,hargrove2015_antiope}. Since asteroid family members are believed to be compositionally similar \citep[e.g.,][]{masiero2015_astfamilies_ast4}, these findings suggest that ice could be widespread within the family.

Interesting dynamical results related to the main asteroid belt have also been reported recently.  \citet{fernandez2015_jfcinterlopers} found that certain known Jupiter-family comets (JFCs) are significantly more dynamically stable than other JFCs, and proposed that they could originate in the asteroid belt, rather than the outer solar system. Meanwhile, \citet{hsieh2016_tisserand} found dynamical pathways by which synthetic test particles with initially asteroid-like orbits could evolve onto JFC-like orbits, and reached a similar conclusion that the JFC population could contain asteroidal interlopers.  However, \citet{fernandez2015_jfcinterlopers} did not trace full dynamical pathways from the asteroid belt to their candidate asteroidal JFC interlopers in their study, while \citet{hsieh2016_tisserand} did not use test particles representing real solar system objects for their study.

The Themis family's proximity to the strong 2:1 mean-motion resonance (MMR) with Jupiter (hereafter, ``the 2:1 MMR'') at 3.278~au makes it likely that objects have been removed from the family over time \citep[e.g.,][]{morbidelli1995_familyresonances}.  Additional sources of instability include the 9:4 MMR with Jupiter (hereafter, ``the 9:4 MMR'') at 3.031~au, which bounds the family on its inner edge, and the 11:5 MMR with Jupiter (hereafter, ``the 11:5 MMR'') at 3.077~au, which intersects the family.  If icy Themis family asteroids are able to reach low-perihelion JFC-like orbits with Tisserand parameter values of $T_J<3$ \citep[cf.][]{kresak1972_tisserand} while retaining at least some near-surface ice, they could become active, and be at least superficially observationally and dynamically indistinguishable from JFCs from the outer solar system. To investigate this possibility, we have conducted numerical integrations to investigate the long-term dynamical behavior of Themis family members.

\section{Experimental Design}\label{section:experimental_design}

We investigated the Themis family by selecting all 4782 members identified by \citet{nesvorny2015_pdsastfam} and, following the method of \citet{hsieh2012_288p}, generating 4 dynamical clones per object using $\sigma$ values of $\sigma_a = 0.001$~au, $\sigma_e = 0.001$, and $\sigma_i = 0.01^{\circ}$ for their semimajor axes, eccentricities, and inclinations, respectively. These $\sigma$ values were chosen to be relatively large (e.g., larger than any of orbital elements' formal uncertainties) in order to characterize the dynamical environment occupied by each Themis family asteroid, help account for chaotic effects, and allow for a rough assessment of the potential effects of collisions that could impart one-time impulses to post-impact bodies. 
We integrated each original object and its clones (i.e., 5 test particles per object, or 23\,910 test particles in total) forward in time for 100 Myr under the gravitational influence of the 7 major planets other than Mercury using the hybrid integrator in the {\tt mercury} $N$-body integration package \citep{chambers1999_mercury}. Non-gravitational forces were not included.  Integrations were performed using the Planetary Science Institute Computing Center's (PSICC) Torque cluster as well as a 64 vCPU Google Compute Engine virtual machine\footnote{\tt https://cloud.google.com/compute/}.

In order to investigate the evolution of our test particles, we recorded each particle's intermediate orbital elements (IOEs) at 1000-year intervals.
%over the entire 100 Myr integration period, or the particle is ejected from the solar system (i.e., $a>100$~au).  
To smooth out short-timescale fluctuations, we computed quasi-mean orbital elements by taking running averages of 100 time steps
%(i.e., averaging over a moving $10^5$-yr window) 
centered on each time step.  We then classified these quasi-mean IOEs as within the Themis family (i.e., based on the osculating orbital element bounds of the current family), outside the Themis family but still dynamically similar to main-belt asteroids (MBAs), or dynamically similar to Jupiter-family comets (JFCs) using the following criteria:
\begin{itemize}
\item{Themis family-like: $3.020~{\rm au} < a_{\rm ioe} < 3.270~{\rm au}$, $0.070 < e_{\rm ioe} < 0.225$, $0.040^{\circ} < i_{\rm ioe} < 3.450^{\circ}$ }
\item{Non-Themis MBA-like: $2.064~{\rm au} < a_{\rm ioe} < 3.277~{\rm au}$, $T_J > 3.05$, $q_{\rm ioe} > (Q_{\rm Mars,max} + 1.5 r_{H,{\rm Mars}})$, $Q_{\rm ioe} < (q_{\rm Jup,min} - 1.5 r_{H,{\rm Jup}})$ \citep[following][]{hsieh2016_tisserand}}
\item{JFC-like: $2.00 < T_J < 3.00$, $q_{\rm ioe} < Q_{\rm Jup,min}$, $Q_{\rm ioe} > (q_{\rm Jup,max} - 1.5 r_{H,{\rm Jup}})$ \citep[following][]{tancredi2014_asteroidcometclassification}}
\end{itemize}
where the Hill radii of Mars and Jupiter are $r_{H,{\rm Mars}}=0.007~{au}$ and $r_{H,{\rm Jup}}=0.355~{\rm au}$, respectively.  Mars's maximum aphelion distance over our integration period is $Q_{\rm Mars,max}=1.714~{\rm au}$, Jupiter's minimum and maximum perihelion distances are $q_{\rm Jup,min}=4.880~{\rm au}$ and $q_{\rm Jup,max}=5.071~{\rm au}$, respectively, and Jupiter's minimum aphelion distance is $Q_{\rm Jup,min}=5.333~{\rm au}$. All of these values are obtained from our integrations themselves, where for reference,
Mars's mean aphelion distance is $Q_{\rm Mars,mean}=1.666~{\rm au}$, and Jupiter's mean perihelion and aphelion distances are $q_{\rm Jup,mean}=4.951~{\rm au}$ and $Q_{\rm Jup,mean}=5.455~{\rm au}$, respectively, based on planetary orbital elements from JPL\footnote{\tt https://ssd.jpl.nasa.gov/?planet\_pos}.
To reduce complexity, we did not compute Minimum Orbit Intersection Distances (MOIDs) as part of classifying orbits as JFC-like as \citet{tancredi2014_asteroidcometclassification} did in order to exclude objects that did not have strong gravitational interactions with Jupiter.  We note however that the {\tt mercury} integration software we used indicated that 99.1\% of particles with JFC-like IOEs experienced actual close encounters with Jupiter (within 3$r_{H,{\rm Jup}}$) and thus can safely be assumed to satisfy the fundamental characteristic of having strong interactions with Jupiter underlying the MOID requirement set by \citet{tancredi2014_asteroidcometclassification} for JFC-like objects.

For reference, we also classify IOEs that meet the following dynamical criteria for Centaurs, long period comets (LPCs), %near-Earth objects (NEOs), 
and dynamically asteroidal near-Earth objects (NEOs):
\begin{itemize}
\item{Centaur-like: $2.00 < T_J$, $q_{\rm ioe} > Q_{\rm Jup,max}$, $q_{\rm ioe} < a_{\rm Ura}$}
\item{LPC-like: $T_J < 2.00$}
%\item{NEO-like: $q_{\rm ioe} < 1.3~{\rm au}$}
\item{NEO-like (dynamically asteroidal): $q_{\rm ioe} < 1.3~{\rm au}$, $T_J > 3.05$}
\end{itemize}
where Jupiter's maximum aphelion distance is $Q_{\rm Jup,min}=5.525~{\rm au}$ and Uranus's mean semimajor axis distance is $a_{\rm Ura}=19.201~{\rm au}$.
Since the primary motivation of this work is to specifically investigate connections between the Themis family and JFCs, however, we will not discuss these types of IOEs in detail in this work.

\begin{figure}[htb!]
\centerline{\includegraphics[width=2.6in]{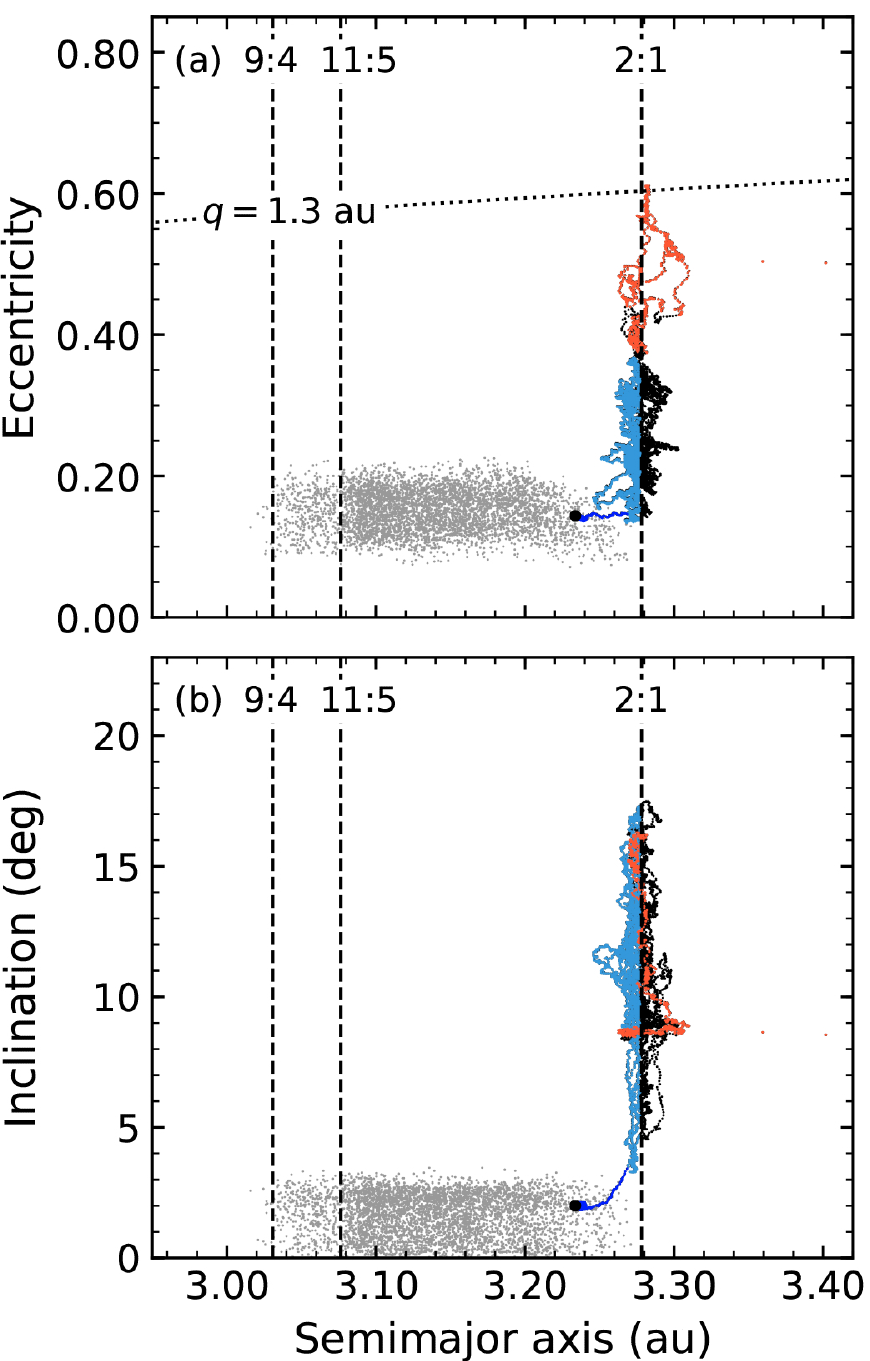}}
\caption{\small Plots of the forward dynamical evolution of a test particle representing Themis family member (12360) Unilandes in (a) semimajor axis versus eccentricity space, and (b) semimajor axis versus inclination space, where large black circles mark the starting orbital elements of the asteroid, small grey dots mark the orbital elements of current Themis family members, small dark blue dots mark IOEs for the asteroid that lie within the osculating orbital element boundaries of current Themis family members, small light blue dots mark IOEs for the asteroid that meet the criteria for main-belt-like orbits described in the text, and small red dots mark IOEs for the asteroid to meet the criteria for JFC-like orbits described in the text.  Vertical dashed lines mark the 9:4, 11:5, and 2:1 mean-motion resonances with Jupiter, as labeled.
}
\label{figure:aei_particle_evolution}
\end{figure}

\begin{figure}[htb!]
\centerline{\includegraphics[width=3.2in]{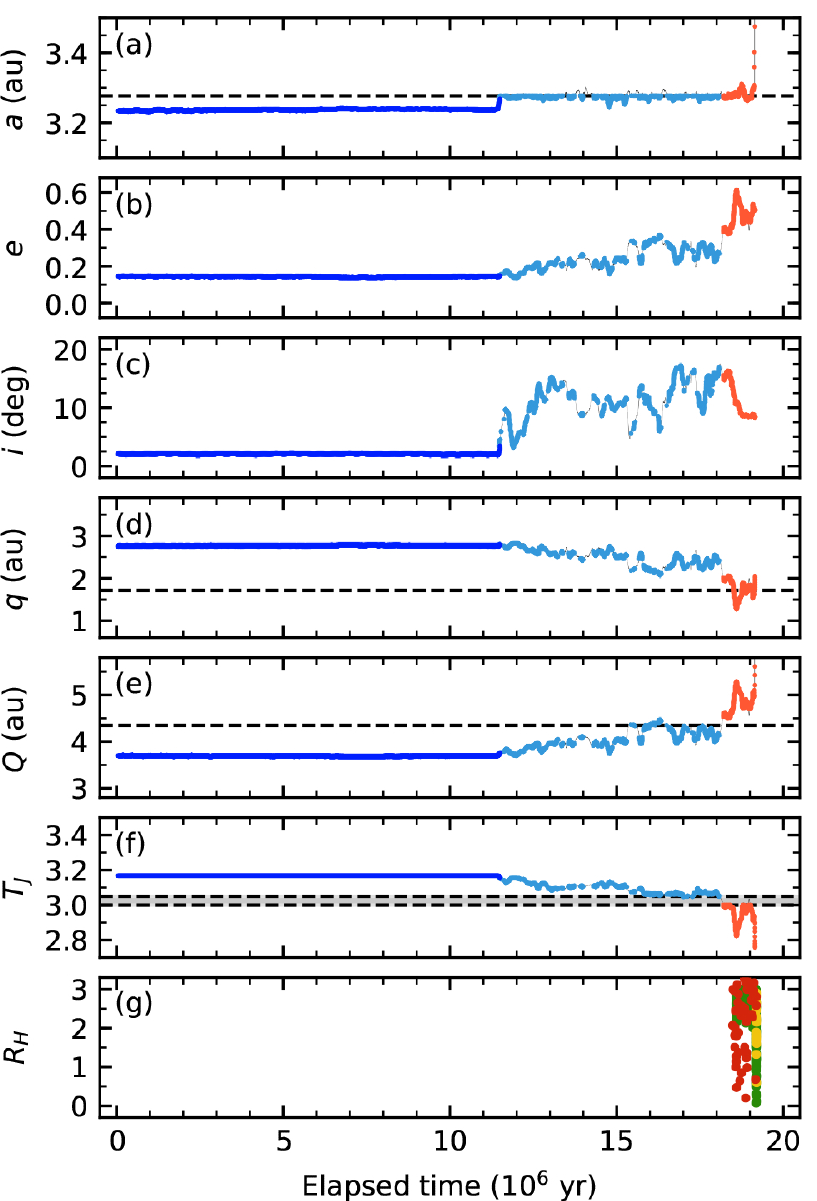}}
\caption{\small Plots of the forward dynamical evolution of a test particle representing Themis family member (12360) Unilandes, specifically (a) semimajor axis (b) eccentricity, (c) inclination, (d) perihelion, (e) aphelion, and (f) $T_J$ versus time, and (g) plots of distances of close encounters with Mars, Saturn, and Jupiter in terms of Hill radii ($R_H$) of each planet as a function of time.  In panels (a) to (d), small dark blue dots mark IOEs for the asteroid that lie within the osculating orbital element boundaries of current Themis family members, small light blue dots mark IOEs for the asteroid that meet the criteria for main-belt-like orbits described in the text, and small red dots mark IOEs for the asteroid to meet the criteria for JFC-like orbits described in the text.  Horizontal dashed lines in panels (a), (d), and (e) mark the heliocentric distances of the 2:1 MMR, $Q_{\rm Mars,max}$, and $q_{\rm Jup,min}$, respectively.  Horizontal dashed lines and the gray shaded region in panel (f) mark the approximate $T_J$ boundary region between dynamically asteroidal and dynamically cometary orbits.  In panel (g), red dots mark close encounters with Mars, green dots mark close encounters with Jupiter, and yellow dots mark close encounters with Saturn.
}
\label{figure:time_particle_evolution}
\end{figure}

To gain a sense of the potential impacts of non-gravitational perturbations on the dynamical evolution of Themis family asteroids as well as perform an independent check of our initial integrations, we also ran two small-scale sets of follow-up integrations including the Yarkovsky effect using the Orbit9 integrator \citep{milani1988_orbit9} in the OrbFit package\footnote{\tt http://adams.dm.unipi.it/orbfit/}.  In the first set of follow-up integrations, we integrated all 4782 catalogued Themis family asteroids under the gravitational influence of the Sun and the 4 major outer planets, and the Yarkovsky effect, but to reduce computational time, did not include the terrestrial planets.  For these integrations, object sizes were determined using catalogued absolute magnitudes and assumed geometric albedos of $p=0.07$.  A reference maximum drift rate of $(da/dt)_{{\rm max,1km}}=6\times10^{-4}$~au~Myr$^{-1}$ for a body with a diameter of $D=1$~km was adopted from \citet{spoto2015_astfamages} and scaled for each object's size using $(da/dt)_{\rm max} = (da/dt)_{{\rm max,1km}} / D$ as the maximum drift rate for objects with diameters of $D$ in km \citep[cf.][]{bottke2006_yarkovsky}.

%Since the Yarkovsky effect scales as 1/D, the particle sizes are used to calculate the corresponding value of $(da/dt)_{\rm max}$ for each test particle, by scaling from the reference value derived for D=1km object.

Assuming an isotropic distribution of spin axis orientations, a random value of $da/dt$ between $-(da/dt)_{\rm max}$ and $(da/dt)_{\rm max}$ was then assigned to each body.
Our second set of follow-up integrations included the gravitational influence of the Sun and 7 major planets (i.e., including the terrestrial planets except for Mercury) and the Yarkovsky effect, but only included 500 test particles ($\sim$10\% of the total Themis family population). Both sets of integrations were run for 100 Myr, i.e., the same duration as our initial integrations. A detailed analysis of these and other integrations to investigate the impacts of non-gravitational effects on Themis family asteroid evolution will be presented in a future paper.  Hereafter, unless otherwise specified, our analysis and discussion will focus on our purely gravitational integrations.

\setlength{\tabcolsep}{4.0pt}
\setlength{\extrarowheight}{0em}
\begin{table*}[htb!]
\centering
\caption{Integration results summary (no Yarkovsky effect)}
\smallskip
\footnotesize
\begin{tabular}{lrcrrccrccrc}
\hline\hline
 & \multicolumn{5}{c}{Entire Themis family$^a$}
 & & \multicolumn{5}{c}{2:1 MMR region$^b$} \\
 & \multicolumn{1}{c}{$n_{\rm real}$$^c$}
 & \multicolumn{1}{c}{$f_{\rm real}$$^d$}
 & & \multicolumn{1}{c}{$n_{\rm all}$$^e$}
 & \multicolumn{1}{c}{$f_{\rm all}$$^f$}
 & & \multicolumn{1}{c}{$n_{\rm real}$$^c$}
 & \multicolumn{1}{c}{$f_{\rm real}$$^d$}
 & & \multicolumn{1}{c}{$n_{\rm all}$$^e$}
 & \multicolumn{1}{c}{$f_{\rm all}$$^f$} \\
\hline
Dynamically stable & 4750 & 0.9933 & & 4415 & 0.9233 & & 680 & 0.9605 & & 421 & 0.5946 \\
~~~~Themis only$^g$ & 4692 & 0.9812 & & 3514 & 0.7348 & & 678 & 0.9576 & & 351 & 0.4958 \\
~~~~Main-belt, non-Themis$^h$ & 58 & 0.0121 & & 901 & 0.1884 & & 2 & 0.0028 & & 70 & 0.0989 \\
\hline
Dynamically unstable & 32 & 0.0067 & & 367 & 0.0767 & & 28 & 0.0395 & & 287 & 0.4054 \\
~~~~JFC-like$^i$ & 32 & 0.0067 & & 365 & 0.0767 & & 28 & 0.0395 & & 287 & 0.4054 \\
~~~~Centaur-like$^j$ & 8 & 0.0017 & & 129 & 0.0270 & & 6 & 0.0085 & & 86 & 0.1215 \\
%~~~~NEO-like (all)$^k$ & 20 & 0.0042 & & 249 & 0.0521 & & 0 & 0.0014 & & 11 & 0.0155 \\
~~~~NEO-like ($T_J>3.05$)$^k$ & 3 & 0.0006 & & 25 & 0.0052 & & 1 & 0.0014 & & 11 & 0.0155 \\
~~~~LPC-like$^l$ & 2 & 0.0004 & & 72 & 0.0151 & & 2 & 0.0028 & & 67 & 0.0946 \\
\hline
Total & 4782 & 1.0000 & & 4782 & 1.0000 & & 708 & 1.0000 & & 708 & 1.0000 \\
\hline
\hline
\multicolumn{12}{l}{\vspace{-0.1cm}$^a$ Results for all 4782 Themis family members} \\
\multicolumn{12}{l}{\vspace{-0.1cm}$^b$ Results for Themis family members with osculating elements within the region of influence of  } \\
\multicolumn{12}{l}{\vspace{-0.1cm}~~~~ the 2:1 MMR described in the text} \\
\multicolumn{12}{l}{\vspace{-0.1cm}$^c$ Number of test particles representing real Themis family members in specified category} \\
\multicolumn{12}{l}{\vspace{-0.1cm}$^d$ Fraction of test particles representing real Themis family members in specified category} \\
\multicolumn{12}{l}{\vspace{-0.1cm}$^e$ Number of Themis family members with one or more test particles representing dynamical } \\
\multicolumn{12}{l}{\vspace{-0.1cm}~~~~ clones in specified category} \\
\multicolumn{12}{l}{\vspace{-0.1cm}$^f$ Fraction of Themis family asteroids with one or more test particles representing dynamical} \\
\multicolumn{12}{l}{\vspace{-0.1cm}~~~~  clones in specified category} \\
\multicolumn{12}{l}{\vspace{-0.1cm}$^g$ Test particles that remain within the Themis family for the entire integration period} \\
\multicolumn{12}{l}{\vspace{-0.1cm}$^h$ Test particles that evolve beyond the Themis family but have MBA-like IOEs for the entire } \\
\multicolumn{12}{l}{\vspace{-0.1cm}~~~~ integration period } \\
\multicolumn{12}{l}{\vspace{-0.1cm}$^i$ Test particles with JFC-like IOEs} \\
\multicolumn{12}{l}{\vspace{-0.1cm}$^j$ Test particles with Centaur-like IOEs} \\
\multicolumn{12}{l}{\vspace{-0.1cm}$^k$ Test particles with $T_J>3$ NEO-like IOEs} \\
\multicolumn{12}{l}{\vspace{-0.1cm}$^l$ Test particles with LPC-like IOEs} \\
\end{tabular} \\
\label{table:integrations_summary}
\end{table*}

\section{RESULTS AND ANALYSIS}\label{section:results}

\subsection{Overview}\label{section:results_overview}

The results of our primary integrations are summarized in Table~\ref{table:integrations_summary}.  We find that the vast majority of Themis family asteroids and their dynamical clones are dynamically stable and remain within the family over our entire 100~Myr integration period, as expected for members of a large low-inclination, moderate-eccentricity asteroid family in the main asteroid belt.  We do however find that 58 real Themis family members (1.2\% of the family) and dynamical clones of an additional 843 family members (17.6\% of the family) had IOEs outside the Themis family but still within the asteroid belt for the entire integration period.  Meanwhile, 32 real Themis family members (0.7\% of the family) and dynamical clones of an additional 335 family members (7.0\% of the family) were ejected from the solar system ($a>100~{\rm au}$) within the integration period.  Virtually all of these unstable test particles had JFC-like IOEs, where a small number of these particles also reached Centaur-like, LPC-like, or dynamically asteroidal NEO-like IOEs.

\begin{figure*}[tbp]
\centerline{\includegraphics[width=6in]{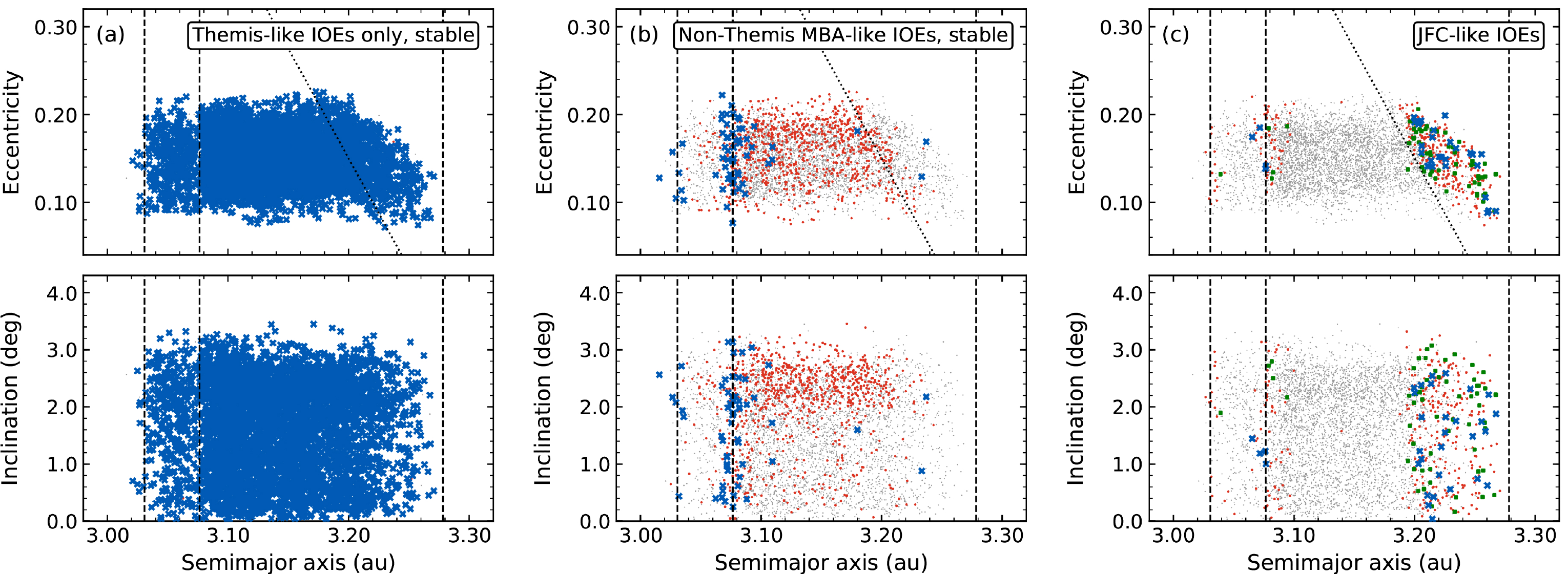}}
\caption{\small Semimajor axis vs. eccentricity (upper panels) and inclination (lower panels) plots showing the starting osculating orbital elements of test particles that (a) remain within the Themis family for the entire integration period, (b) evolve beyond the Themis family but remain within the main belt for the entire integration period, and (c) reach JFC-like IOEs during the integration period.  In all panels, small gray dots mark the orbital elements of all currently known Themis family members, blue X's mark real Themis family asteroids, and small red dots mark dynamical clones, while in panel (c), green squares mark Themis family members found to reach JFC-like IOEs in our first set of follow-up integrations including the Yarkovsky effect. Vertical dashed lines in each panel denote, from left to right, the 9:4, 11:5, and 2:1 mean-motion resonances with Jupiter.  Diagonal dotted lines in each panel show the boundary of the empirically identified region of influence of the 2:1 MMR in semimajor axis versus eccentricity space.}
\label{figure:aei_dynamical_behavior}
\end{figure*}

\begin{figure*}[tbp]
\centerline{\includegraphics[width=6.5in]{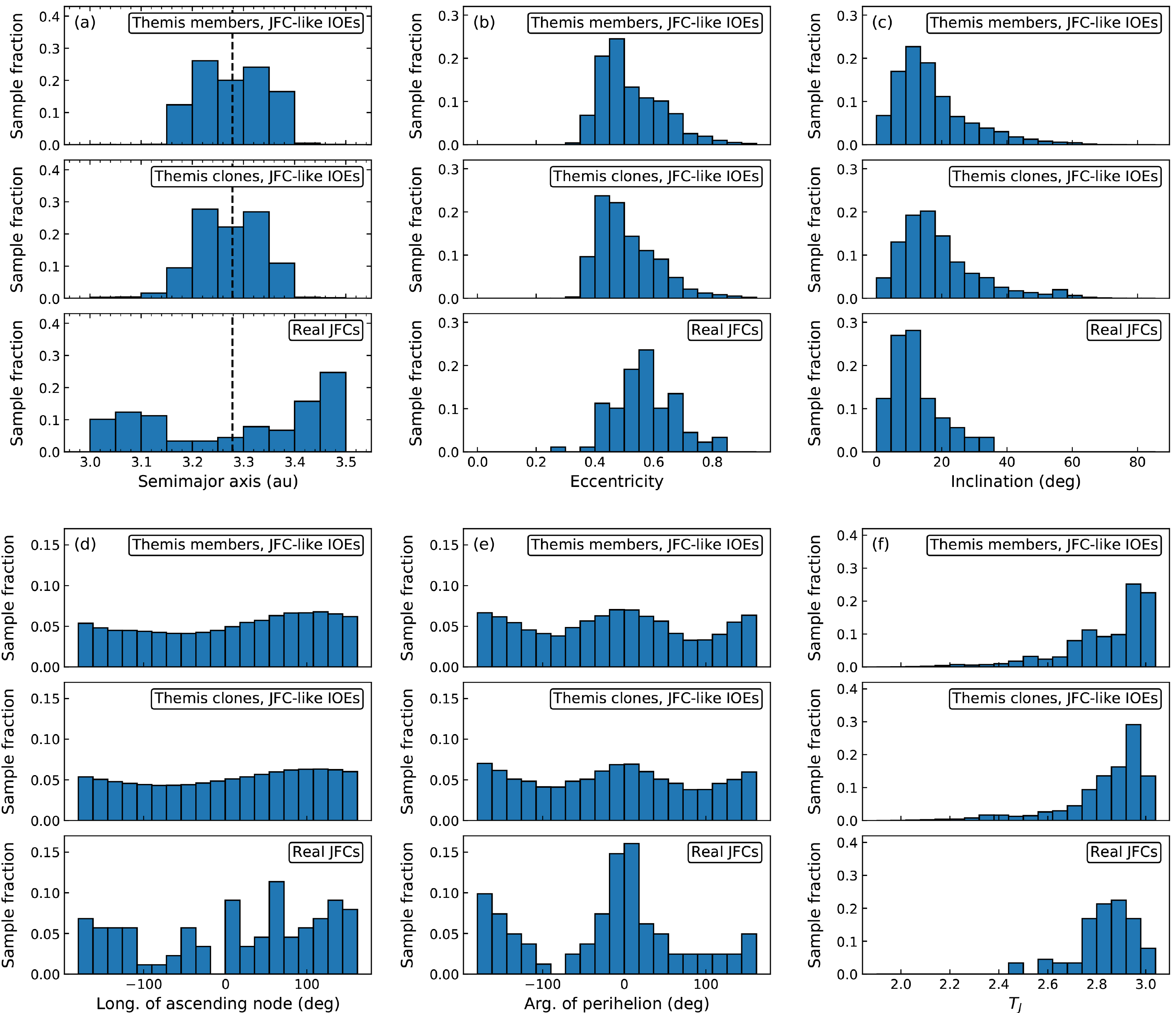}}
\caption{\small Histograms of the (a) semimajor axes (where the location of the 2:1 MMR is marked with a vertical dashed line in all panels), (b) eccentricities, (c) inclinations, (d) longitudes of the ascending node, (e) arguments of perihelion, and (f) Tisserand parameters with respect to Jupiter of the JFC-like IOEs of real Themis family objects (top panels), the JFC-like IOEs of dynamical clones (middle panels), and current orbits of real JFCs (bottom panels), restricted to only those JFC-like IOEs of test particles and the 89 real JFCs where $3.0~\rm{au} < a_{\rm osc} < 3.5~\rm{au}$.}
\label{figure:aei_jfc-like_orbital_elements}
\end{figure*}

In Figures~\ref{figure:aei_particle_evolution} and \ref{figure:time_particle_evolution}, we show an example of the dynamical evolution undergone by a test particle representing current Themis family member (12360) Unilandes that was found to evolve from a Themis-family-like orbit to a JFC-like orbit during our integrations.
As with many of the clones found to reach JFC-like orbits (as defined in Section~\ref{section:experimental_design}) in our integrations, this particle starts out in the region affected by the 2:1 MMR with Jupiter (cf.\ Figure~\ref{figure:aei_particle_evolution}). The resonance then causes eccentricity and inclination excitation (Figure~\ref{figure:time_particle_evolution}b,c), eventually driving the particle to Mars- and Jupiter-crossing orbits (Figure~\ref{figure:time_particle_evolution}d,e) and comet-like $T_J$ values (Figure~\ref{figure:time_particle_evolution}f) as its increasing eccentricity causes it to interact more strongly with the major planets (Figure~\ref{figure:time_particle_evolution}g).

Semimajor axis versus eccentricity and inclination plots (Figure~\ref{figure:aei_dynamical_behavior}) indicate that the vast majority of real Themis family asteroids with IOEs outside the Themis family but still within the asteroid belt largely originate near the 11:5 MMR.  Meanwhile, although a few test particles with JFC-like IOEs also originate near the 11:5 MMR, the vast majority start near the 2:1 MMR.  We mark the boundary of the region affected by the 2:1 MMR in Figure~\ref{figure:aei_dynamical_behavior}c.  This boundary is visually approximated from the extent in semimajor axis-eccentricity space of the starting orbital elements of test particles found to eventually evolve onto JFC-like orbits, and is given by $e=-2.5a + 8.15$ (for $a$ in au).
We use an empirically approximated boundary for the 2:1 MMR's region of influence rather than a theoretically derived resonance boundary \citep[e.g.,][]{murray2000_solarsystemdynamics,wang2017_mmrs}, because the former is simpler to implement and the potentially greater accuracy of the latter is not necessary for our purposes in this work.

%We use an empirically approximated boundary for the 2:1 MMR's region of influence rather than a theoretically derived resonance boundary \citep[e.g.,][]{murray2000_solarsystemdynamics,wang2017_mmrs}, as these are not necessarily more precise in situations such as this one since they typically assume a simplified three-body planar framework for their derivations which only approximates the real solar system. Our objective here is also simply to identify the approximate region of the Themis family from which most objects that evolve onto JFC-like orbits originate, and so for this purpose, an empirically approximated boundary for the 2:1 MMR's region of influence based on our ``real-world'' integration results is more suitable than a theoretically derived one.
For reference, there are 708 known Themis family members (Table~\ref{table:integrations_summary}) and $\sim$13\,000 total main-belt asteroids in the near-MMR region of interest that we have identified.  Of those 708 known Themis family members ($\sim$15\% of the family), 28 real family members (4.0\% of this subset of the family) and dynamical clones of an additional 259 family members (36.6\% of this subset of the family) reach JFC-like IOEs during our integrations.

In our first set of follow-up integrations including the Yarkovsky effect, we find that even more test particles reach JFC-like orbits than in our purely gravitational integrations.  Specifically, we find that 73 Themis family asteroids, 67 of which are located in the region of influence of the 2:1 MMR discussed above, reach JFC-like IOEs, representing roughly a factor of two increase over the number of Themis family asteroids with JFC-like IOEs in our original integrations. In our second set of follow-up integrations (including the Yarkovsky effect and all seven major planets besides Mercury as perturbers), we find that 8 out of 500 particles reach JFC-like IOEs, representing an additional $\sim$5-10\% increase over the number of particles with JFC-like IOEs from our first set of follow-up integrations.

%In our presentation, we will report on our findings in greater detail, discussing follow-up efforts to confirm these early results, estimates of the rate at which Themis family objects may escape the main belt and the survivability of ice in these objects during the journey to and while in near-Earth space.

%%%%%%%%%%%%%%%%%%%%%%%%%%%%%%%%%%%%%%%%%%%%%%%%%%%%%%%%%%%%%%%%%%%%%%%%%%%%%%%%%%%%%%%%%%%%%%

\subsection{Analysis of JFC-like particles}\label{section:results_jfclike_particles}

We seek to better assess whether icy objects originating in the Themis family could plausibly be mistaken for JFCs from the outer solar system, and if so, how they might potentially be identified.  To do so, we examine the distribution in orbital element space of the JFC-like IOEs attained by particles in our integrations and compare it to those of objects in the present-day known JFC population.

One immediately apparent feature of the JFC-like IOEs from our integrations is that 88.8\% have semimajor axes within 0.125~au of the 2:1 MMR (i.e., between 3.15~au and 3.40~au), likely reflecting the strong role that this resonance plays in driving these particles onto JFC-like orbits.  Focusing on this particular region, we compare the semimajor axis, eccentricity, inclination, longitude of the ascending node, argument of perihelion, and $T_J$ distributions of the 790\,765 IOEs of test particles representing both real Themis family asteroids and their clones with semimajor axes between 3.0~au and 3.5~au with the 89 real JFCs catalogued by JPL\footnote{\tt https://ssd.jpl.nasa.gov/?sb\_elem} as of 2019 October 1 found within the same semimajor axis range (Figure~\ref{figure:aei_jfc-like_orbital_elements}).

The semimajor axis distribution of the sample of real JFCs plotted in Figure~\ref{figure:aei_jfc-like_orbital_elements} shows that there are actually relatively few JFCs in the semimajor axis range within which the majority of the JFC-like IOEs of our test particles are found, suggesting that if dynamically evolved Themis family asteroids are indeed present among active JFCs, they may only comprise a small subset of the population.  As such, close matches of the orbital element distributions of the JFC-like IOEs of our test particles and the real JFC population are not necessarily expected or required to validate our results, as long as they are not fundamentally incompatible with each other.  The distributions of the longitudes of the ascending node and arguments of perihelion for our JFC-like test particles do differ somewhat from those of real JFCs with $3.0~{\rm au}<a<3.5~{\rm au}$, but given the natural circulation of these angles over time and the far smaller sample of real JFCs relative to JFC-like test particle IOEs, we do not regard the differences in these distributions as significant. Meanwhile, while the JFC-like IOEs of our test particles cover a somewhat larger inclination range than those real JFCs, we otherwise find that those IOEs span similar ranges of eccentricities, inclinations, and $T_J$ values as real JFCs within the targeted semimajor axis range and have similar distributions within those ranges, and are therefore generally compatible with the real JFC population.

%This result indicates that during the time that these particles meet the criteria of having JFC-like orbital elements defined in Section~\ref{section:experimental_design}, they should be able to effectively mimic other JFCs from the outer solar system if only their orbital elements are considered.

%For reference, 3.3\% have semimajor axes smaller than 3.15~au and 7.9\% have semimajor axes larger than 3.40~au.

\begin{figure}[htb!]
\centerline{\includegraphics[width=3.0in]{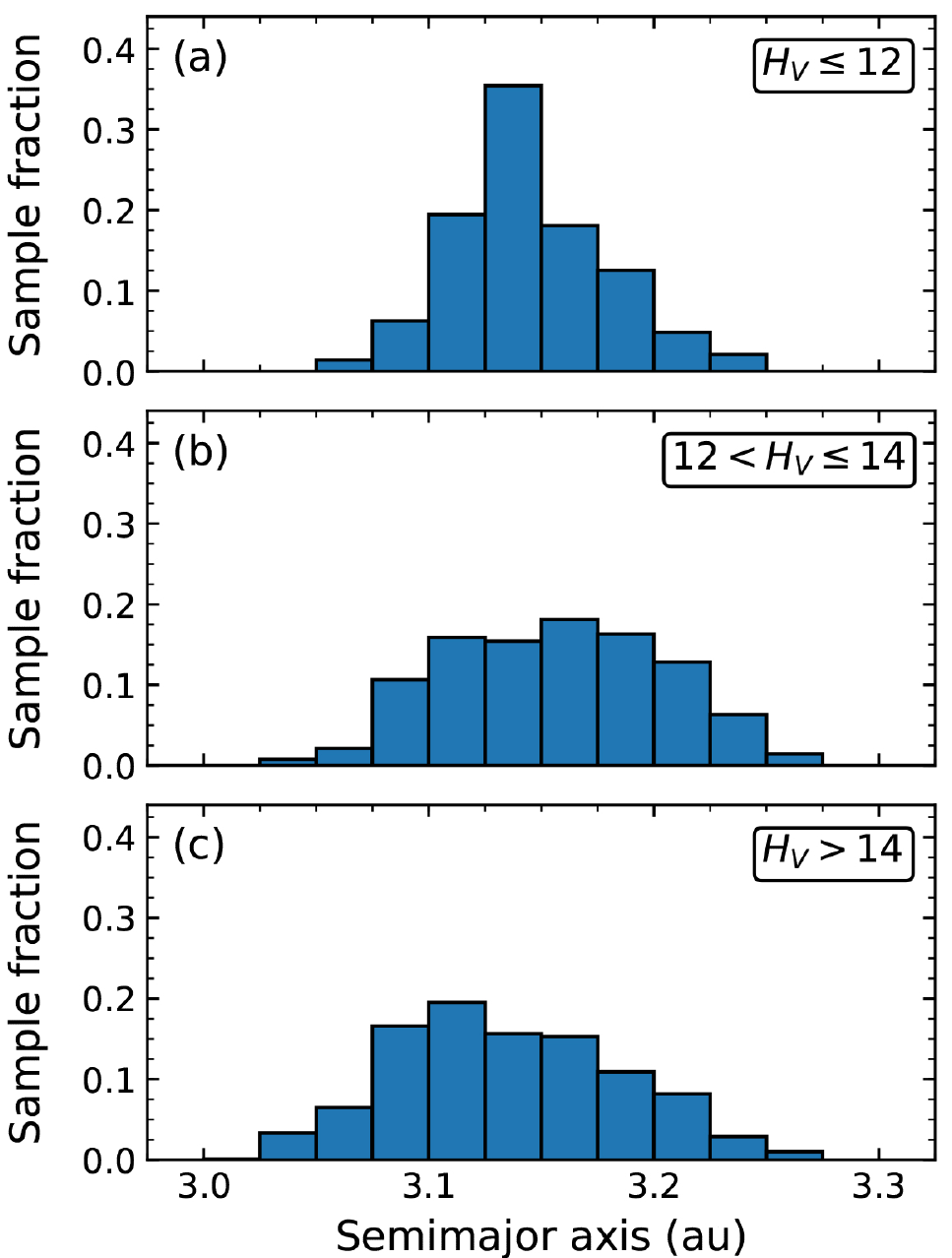}}
\caption{\small Histograms of semimajor axis distributions of Themis family asteroids in three different absolute magnitude bins --- (a) $H_V\leq12$, (b) $12 < H_V \leq 14$, and (c) $H_V > 14$. --- where the population of currently known outer main belt asteroids is believed to be complete for $H_V < 14.4$ \citep[cf.][]{granvik2017_asteroidescape}, meaning that the distributions in panels (a) and (b) should reflect complete sub-populations.
}
\label{figure:themis_histograms}
\end{figure}

\begin{figure}[htb!]
\centerline{\includegraphics[width=3.0in]{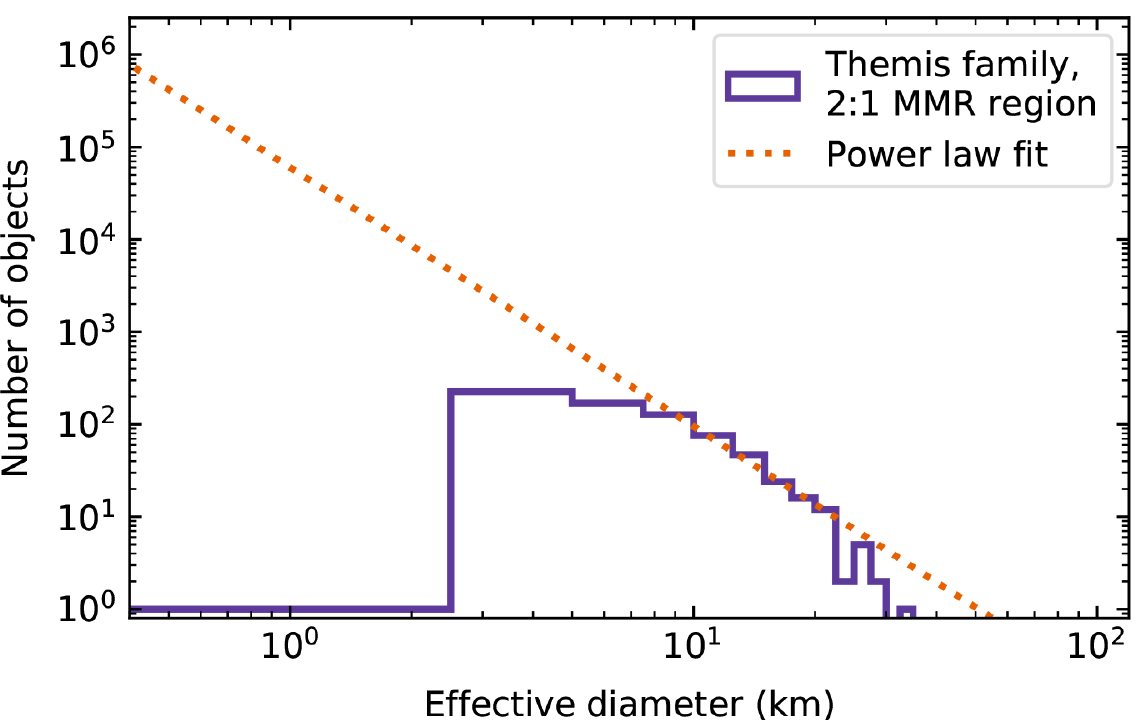}}
\caption{\small Differential size-frequency distributions of the portion of the Themis family found within the region of influence of the 2:1 MMR defined in Section~\ref{section:results_overview} (solid purple line) along with a corresponding power law fit (dotted orange line). 
}
\label{figure:themis_sfd}
\end{figure}

To estimate the potential contribution of Themis family asteroids to the JFC population based on the integrations presented here, 
%\addedtext{we first perform a brief assessment of how representative the Themis family population used for our integrations is of the total population.  \citet{granvik2017_asteroidescape} estimate that known population of main-belt asteroids in the outer main belt (beyond the 3:1 MMR with Jupiter at 2.501~au) is largely complete for $H<14.4$, corresponding to $D_{\rm eff}\sim8$~km.  We plot histograms (Figure~\ref{figure:themis_histograms}) of the osculating semimajor axis, eccentricity, and inclination distributions of the Themis family objects used in our integrations, divided into three separate absolute magnitude categories: $H_V \leq 12$ (144 objects), $12 < H_V \leq 14$ (882 objects), and $H_V > 14$ (3756 objects).  The populations of the first two absolute magnitude categories should be complete according to \citet{granvik2017_asteroidescape}, yet there are notable differences
%}
%\addedtext{}
we note that measurements of 89 JFC nuclei \citep{fernandez2013_jfcnuclei} found nuclei diameters down to $\sim$1~km.  We convert absolute magnitudes ($H_V$) catalogued by the Minor Planet Center for Themis family asteroids to effective diameters ($D_{\rm eff}$) in km using
\begin{equation}
D_{\rm eff} = \left({8.96\times10^{16}\over p_V} \times 10^{0.4(m_{\odot} - H_V)}\right)^{0.5}
\end{equation}
assuming $V$-band albedos of $p_V=0.05$ and a $V$-band solar magnitude of $m_{\odot}=-26.71$~mag \citep[cf.][]{jewitt1991_cometphotometry_cometsposthalley,fernandez2000_encke}.  
By fitting a power law to the differential size-frequency distribution of the source population of the JFC-like particles in our integrations and extrapolating to smaller sizes where the known population is likely incomplete, we should be able to use the fraction of particles that reach JFC-like orbits in our integrations and the total length of our integration period to estimate the total flux of objects onto such orbits.

We note, however, that plots of the semimajor axis distributions of Themis family asteroids in different size bins (Figure~\ref{figure:themis_histograms}) show that larger asteroids are more concentrated toward the center of the family in semimajor axis space, with a larger fraction of smaller asteroids found farther from the center. This size-dependent semimajor axis distribution is consistent with the previously noted phenomenon of size-dependent Yarkovsky-driven semimajor axis drift in asteroid families \citep[e.g.,][]{vokrouhlicky2006_yarkovskyfamilies,bottke2006_yarkovsky,walsh2013_eulaliapolana}.  To mitigate the effect of this uneven distribution of objects of different sizes on our results, we restrict our analysis on the region of influence of the 2:1 MMR defined in Section~\ref{section:results_overview}, instead of the full family.

%\addedtext{Restricting the sample to just the region of influence of the 2:1 MMR defined in Section~\ref{section:results_overview}
%Since the majority of the Themis family objects found to reach JFC-like orbits originate in the region of influence of the 2:1 MMR defined in Section~\ref{section:results_overview}, it would be more instructive to estimate the size of the potential source population in this region alone.}

Visually fitting a power law to the differential size-frequency distribution of Themis family asteroids in the region of influence of the 2:1 MMR (Figure~\ref{figure:themis_sfd}), we find an approximate slope parameter of $-2.8$ and an estimated $\sim5\times10^4$ Themis family asteroids with diameters of 1~km in this region.
Assuming that the fraction ($\sim$4\%; cf.\ Table~\ref{table:integrations_summary}) of known Themis family members in the near-MMR region that reach JFC-like orbits over a 100 Myr integration period found in our purely gravitational integrations remains similar for the complete population in this region, we expect on the order of $\sim$2000 Themis family objects from the near-MMR region could evolve onto JFC-like orbits every 100 Myr under the influence of gravity alone.  
We find a mean residence time on JFC-like orbits of $t\sim1\times10^6$~yr for Themis family objects that reach such orbits during our integrations. Thus, the $\sim$2000 objects per 100 Myr found to reach JFC-like orbits should collectively spend a total of $\sim2\times10^9$~yr on such orbits, suggesting that, on average, we should expect $\sim$20 objects from the Themis family on JFC-like orbits at any given time with the potential to mimic active JFCs from the outer solar system.
%)\citep[cf.][]{schorghofer2018_asteroidiceloss}}.

There are many uncertainties associated with this analysis, however.  First, if the ``wavy'' shape of the main asteroid belt's overall size frequency distribution (which tends toward a shallower slope near $D\sim1$~km) noted by \citet{bottke2005_sizedistribution} applies to the Themis family, our simple power-law extrapolation of the observed size distribution of asteroids in the near-MMR region may overestimate the number of $D\geq1$~km Themis family asteroids by a factor of a few.  That said, \citet{bottke2005_sizedistribution} also specifically model a ``Themis-style'' catastrophic fragmentation event and find a size frequency distribution that is relatively well characterized by a single power law over the range of sizes ($D\sim1$~km to $D\sim20$~km) that is relevant to our analysis. It is also possible that some nuclei of observably active comets could be smaller than $D=1~{\rm km}$.  Pushing to smaller sizes would increase the size of the relevant source population in the Themis family, which would then proportionally increase the expected number of relevant objects from the family that escape onto JFC-like orbits.

As discussed earlier, non-gravitational forces may increase the expected rate of Themis family objects evolving onto JFC-like orbits compared to the rate computed from purely gravitational integration results by a factor of two or more, while the much higher fraction of Themis family asteroids with dynamical clones relative to just the objects themselves that evolve onto JFC-like orbits indicates that other perturbations (e.g., collisions) could further increase this flux.  Lastly, we see indications that the 100~Myr integration period we use in this work may be too short to capture the full evolution of Themis family objects evolving via resonances other than the 2:1 MMR, potentially causing our analysis to underestimate the full steady-state flux of objects that ultimately reach JFC-like orbits (Section \ref{subsection:jfcpopulation}).

Given all of these complicating factors, our calculation above that $\sim$20 objects from the Themis family with the potential to mimic active JFCs from the outer solar system should be present at any given time should be considered an order of magnitude estimate only.  As such, we conclude that there may be tens of such objects at any given time (compared to a total population of $\sim$600 known JFCs catalogued by JPL\footnote{\tt https://ssd.jpl.nasa.gov/?sb\_elem} as of 2019 October 1).

%%%%%%%%%%%%%%%%%%%%%%%%%%%%%%%%%%%%%%%%%%%%%%%%%%%%%%%%%%%%%%%%%%%%%%%%%%%%%%%%%%%%%%%%%%%%%%

\section{Discussion}\label{section:discussion}

\subsection{Possible dynamical pathways}\label{subsection:pathway}

Investigation of the evolution of individual particles that reach JFC-like IOEs in our purely gravitational integrations indicate that most of these particles experience eccentricity excitation due to the nearby 2:1 MMR.  This excitation in turn lowers their perihelia, causing them to interact more strongly with planets other than Jupiter, with which we find that many test particles experience numerous close encounters during their transition to low-$T_J$ JFC-like orbits.  
We specifically find that 58.8\%, 84.6\%, 99.7\%, and 87.7\% of real Themis family asteroids and their dynamical clones with JFC-like IOEs experience close encounters with Venus, Earth, Mars, and Saturn, respectively.  These interactions likely play a significant role in driving these particles to comet-like $T_J$ values considering that these other planets represent additional massive perturbers that are not accounted for in the idealized 3-body system (i.e., the Sun, Jupiter, and the small body in question) on which the $T_J$ approximation is based.  

In our first set of follow-up integrations which included the Yarkovsky effect but no terrestrial planets, encounters with Venus, Earth, and Mars of course did not occur, but evolution onto low-$T_J$ JFC-like orbits by Themis family asteroids still did. This suggests that the observed evolution of particles towards comet-like $T_J$ values can also be driven by a combination of the Yarkovsky effect and interactions with Saturn alone as an additional planetary perturber, even for objects that do not have close encounters with the terrestrial planets.

%%%%%%%%%%%%%%%%%%%%%%%%%%%%%%%%%%%%%%%%%%%%%%%%%%%%%%%%%%%%%%%%%%%%%%%%%%%%%%%%%%%%%%%%%%%%%%

\subsection{Implications for currently known JFCs}\label{subsection:knownjfcs}

If Themis family objects are present on JFC-like orbits in appreciable numbers, the next key question is whether they would be expected to be icy enough to support sublimation like other JFCs.  Expected depths-to-ice in thermal models of asteroids depend on factors like the thermal properties of mantle material, obliquity, and latitude, but these depths have the potential to be quite shallow at the polar regions of low-obliquity outer main-belt objects formed in recent catastrophic disruption events \citep{schorghofer2008_mbaice,schorghofer2016_asteroidice}. Thermal modeling by \citet{schorghofer2018_asteroidiceloss} indicates that young (e.g., $\lesssim10$~Myr old), icy outer-belt asteroids could still retain polar near-surface ice under certain conditions, and therefore could potentially join the active JFC population for a (currently poorly-constrained) period of time.

%\addedtext{That said, whether Themis family objects that evolve onto JFC-like orbits actually become active  and how long they remain active depends on several unknown factors such as each body's initial ice content and ice retention history over its lifetime \citep[cf.][]{schorghofer2018_asteroidiceloss}.  As such, we \addedtext{currently} cannot meaningfully estimate the fraction of the population of dynamically JFC-like objects with Themis family origins that that should be active at any given time, and therefore cannot yet estimate the number of past Themis family asteroids that may be infiltrating the currently active JFC population today.
%(i.e., actually mimicking JFCs when observed and not simply being among the $\sim$500 currently known inactive asteroids with JFC-like orbits within our semimajor axis range of interest)
%}

\setlength{\tabcolsep}{4.0pt}
\setlength{\extrarowheight}{0em}
\begin{table*}[htb!]
\centering
\caption{JFCs with $3.15~{\rm au}<a<3.40~{\rm au}$}
\smallskip
\footnotesize
\begin{tabular}{lrcrrcc}
\hline\hline
\multicolumn{1}{c}{Comet}
 & \multicolumn{1}{c}{$a$$^a$}
 & \multicolumn{1}{c}{$e$$^b$}
 & \multicolumn{1}{c}{$i$$^c$}
 & \multicolumn{1}{c}{$q$$^d$}
 & \multicolumn{1}{c}{$Q$$^e$}
 & \multicolumn{1}{c}{$T_J$$^f$} \\
\hline
 16P/Brooks 2               & 3.358 & 0.563 &  4.26 & 1.4667 & 5.250 & 2.873 \\
 43P/Wolf-Harrington        & 3.350 & 0.595 & 15.97 & 1.3570 & 5.343 & 2.793 \\
83D/Russell 1               & 3.338 & 0.517 & 22.66 & 1.6115 & 5.064 & 2.824 \\
104P/Kowal 2                & 3.263 & 0.639 & 10.25 & 1.1791 & 5.347 & 2.794 \\
124P/Mrkos                  & 3.309 & 0.506 & 31.72 & 1.6348 & 4.984 & 2.742 \\
210P/Christensen            & 3.193 & 0.829 & 10.18 & 0.5445 & 5.842 & 2.491 \\
213P/Van Ness               & 3.347 & 0.407 & 10.38 & 1.9833 & 4.711 & 2.995 \\
218P/LINEAR                 & 3.342 & 0.491 & 18.17 & 1.7011 & 4.983 & 2.884 \\
224P/LINEAR-NEAT            & 3.342 & 0.437 & 14.73 & 1.8818 & 4.803 & 2.951 \\
267P/LONEOS                 & 3.290 & 0.593 &  5.37 & 1.3377 & 5.242 & 2.856 \\
294P/LINEAR                 & 3.200 & 0.595 & 19.09 & 1.2977 & 5.103 & 2.818 \\
304P/Ory                    & 3.243 & 0.574 &  2.76 & 1.3822 & 5.104 & 2.896 \\
337P/WISE                   & 3.291 & 0.498 & 15.40 & 1.6530 & 4.929 & 2.911 \\
365P/PANSTARRS              & 3.186 & 0.573 &  9.84 & 1.3591 & 5.012 & 2.897 \\
D/1770 L1 (Lexell)          & 3.153 & 0.786 &  1.55 & 0.6744 & 5.632 & 2.612 \\
D/1978 R1 (Haneda-Campos)   & 3.290 & 0.665 &  5.95 & 1.1014 & 5.479 & 2.762 \\
P/2000 R2 (LINEAR)          & 3.338 & 0.584 &  3.22 & 1.3899 & 5.287 & 2.857 \\
P/2008 WZ96 (LINEAR)        & 3.357 & 0.510 &  6.96 & 1.6462 & 5.067 & 2.922 \\
P/2013 T2 (Schwartz)        & 3.393 & 0.528 &  9.35 & 1.5996 & 5.185 & 2.886 \\
P/2013 YG46 (Spacewatch)    & 3.305 & 0.454 &  8.11 & 1.8044 & 4.806 & 2.980 \\
P/2015 J3 (NEOWISE)         & 3.350 & 0.554 &  8.13 & 1.4940 & 5.207 & 2.876 \\
P/2016 P1 (PANSTARRS)       & 3.230 & 0.294 & 25.77 & 2.2805 & 4.179 & 2.967 \\
P/2019 A8 (PANSTARRS)       & 3.354 & 0.439 &  2.97 & 1.8828 & 4.826 & 2.992 \\
\hline
\hline
\multicolumn{7}{l}{\vspace{-0.1cm}$^a$ Semimajor axis, in au} \\
\multicolumn{7}{l}{\vspace{-0.1cm}$^b$ Eccentricity} \\
\multicolumn{7}{l}{\vspace{-0.1cm}$^c$ Inclination, in degrees} \\
\multicolumn{7}{l}{\vspace{-0.1cm}$^d$ Perihelion distance, in au} \\
\multicolumn{7}{l}{\vspace{-0.1cm}$^e$ Aphelion distance, in au} \\
\multicolumn{7}{l}{\vspace{-0.1cm}$^f$ Tisserand parameter value with respect to Jupiter} \\
\multicolumn{7}{l}{\vspace{-0.1cm}$^g$ References: [1] \citep{ahearn1995_ensemblecomets}} \\
\end{tabular} \\
\label{table:real_jfcs}
\end{table*}

As of 2019 November 1, there are 23 known active JFCs (Table~\ref{table:real_jfcs}; where three are classified as defunct) with $3.15~{\rm au}<a<3.40~{\rm au}$, the semimajor axis range containing most of the JFC-like IOEs we find for Themis family objects in our integrations.  For reference, an additional $\sim$500 currently inactive asteroids with dynamically JFC-like orbits within this semimajor axis range are also catalogued by JPL.  Notably, 210P/Christensen is the only active JFC among these objects that overlaps with the list of JFCs with potential asteroidal origins identified by \citet{fernandez2015_jfcinterlopers}.
Given that we cannot say how many of the Themis family objects found to evolve onto JFC-like orbits in our integrations will actually become active and for how long from our dynamical results alone (Section~\ref{section:results_jfclike_particles}), we emphasize that we cannot currently say whether any known JFCs may actually be from the Themis family or how many active JFCs with Themis family origins that we should expect to find in general at any given time.

Several of the aforementioned JFCs in our semimajor axis range of interest have been previously observationally characterized, such as 16P/Brooks 2 \citep[$a=3.358$~au; e.g.,][]{sekanina1997_splitcomets,kiselev2002_cometpolarimetry}, 43P/Wolf-Harrington \citep[$a=3.350$~au; e.g.,][and references within]{snodgrass2006_cometphotometry,fink2009_cometsurvey}, 104P/Kowal 2 \citep[$a=3.263$~au; e.g.,][]{snodgrass2006_cometphotometry}, and 124P/Mrkos \citep[$a=3.309$~au; e.g.,][]{licandro2003_124p}, but many others have not been studied in much detail or at all.
Interestingly, 16P/Brooks 2 and 43P/Wolf-Harrington have both been found to be depleted in carbon-chain species \citep[i.e., C$_2$ and C$_3$;][]{schleicher1993_43pwolfharrington,ahearn1995_ensemblecomets,cochran2012_cometspectroscopy}.  Whether such depletion points to possible asteroidal origins is unclear (these two objects are certainly not the only carbon chain depleted comets to be identified), but could warrant further investigation.  Meanwhile, the nucleus of 124P/Mrkos has been found to be spectroscopically similar to D-type asteroids \citep{licandro2003_124p}, which is consistent with spectra obtained for nuclei of other JFCs \citep[e.g.,][]{lamy2004_cometnuclei_comets2}, but not of Themis family asteroids \citep[cf.][]{hsieh2018_activeastfamilies}.  Further systematic physical and compositional characterization of comets with semimajor axes in the range where we expect most of the Themis family objects that evolve onto JFC-like orbits to be found will be very important for assessing whether physical evidence supports our conclusion that Themis family objects may be present in the active JFC population.

%{\bf 16P/Brooks 2 is a comet that split due to tidal interactions from Jupiter from a close encounter within 2.0 Jovian radii in July 1886.}

\subsection{Expected asteroidal contribution to the JFC population}\label{subsection:jfcpopulation}

Some theoretical analyses have found that assumed source populations in the outer solar system (primarily, the scattered disk) do not produce a sufficient number of JFCs to supply the observed population \citep[e.g.,][]{volk2008_jfcsource,nesvorny2017_spcorigin}, suggesting that various model assumptions could require modification, or that other sources may also contribute to the population.
In fact, other studies have shown that JFCs could potentially also originate in the Jovian (or even Neptunian) Trojan and Hilda populations \citep[e.g.,][]{levison1997_trojanevolution,horner2010_trojanjfcs,disisto2005_hildajfcs,disisto2019_trojanevolution}.  Asteroids throughout the main belt are well-known to be driven by MMRs and secular resonances into occasionally escaping from the main-belt and evolving onto NEO orbits \citep[e.g.,][]{bottke2002_neodistribution,michtchenko2016_mbadynamics,greenstreet2012_neodistribution,granvik2017_asteroidescape,granvik2018_neodistribution}. Some of these objects may even reach cometary orbits, though this appears to have generally not been explicitly considered in studies to date investigating NEO origins.

%\addedtext{xxx papers discussing other sources of JFCs like Trojans and Hildas xxx}
The results presented here make it tempting to speculate that the Themis family could be one of the additional sources needed to supply the observed population of JFCs.  However, the narrow semimajor axis range near the 2:1 MMR (which is much narrower than the range over which most JFCs are found; Figure~\ref{figure:aei_jfc-like_orbital_elements}), within which we find most JFC-like objects produced by the Themis family, suggests that the family is, at best, a minor contributor.  That said, this work focuses on current Themis family members, rather than the entire outer asteroid belt, much of which could be intrinsically icy and could also include Themis family members that have escaped the family in the past but currently persist on main-belt orbits.

An additional nuance is that we see indications that a 100 Myr integration period may not be long enough to capture the full steady-state flux of all Themis family asteroids onto JFC-like orbits. In particular, the evolution of particles evolving to JFC-like orbits via resonances other than the 2:1 MMR may not be fully captured by the integrations presented here, which could affect the total number and semimajor axis distribution of JFC-like objects found to be produced by the family.
We recall that a typical pathway followed by a Themis family asteroid that evolves onto a JFC-like orbit involves (1) some initial amount of time spent within the Themis family, (2) eccentricity and inclination excitation due to a nearby MMR, during which the object exits the Themis family but is still confined to the main belt, and (3) eventual evolution onto a JFC-like orbit (Section~\ref{subsection:pathway}).  In our integrations, the median time spent in the second stage of this evolution by test particles that eventually reach JFC-like orbits is 5.1 Myr for particles evolving via the 2:1 MMR, and 42.0 Myr for particles evolving via the 9:4 or 11:5 MMRs. The substantial fraction of our total integration period represented by the latter timescale suggests that there may be many particles that will eventually reach JFC-like orbits that are still in this second stage of their evolution at the end of our integrations (e.g., particles that first reach non-Themis MBA-like orbits relatively late in our integrations or that spend longer than the median amount of time in this stage of their evolution).  %In other words, it is possible that some of the particles found to escape the Themis family but remain within the main asteroid belt over our 100~Myr integrations might actually be found to ultimately reach JFC-like orbits if followed over longer integration periods.
A complete accounting of the steady-state flux of Themis family asteroids to JFC-like orbits should include the eventual contribution of these objects.  A follow-up study using a longer integration period that better accounts for the slower dynamical evolution of objects driven by relatively weak resonances would be very useful for ascertaining both the true steady-state flux and orbital element distribution of Themis family asteroids evolving onto JFC-like orbits.

\subsection{Other future work}\label{subsection:futurework}

Larger-scale dynamical studies that sample the entire outer asteroid belt and use longer integration periods, while also including non-gravitational forces and the terrestrial planets as perturbers, are likely needed to assess the asteroid belt's true contribution to the population of dynamically JFC-like small solar system bodies that could potentially comprise part of the active JFC population if they also happen to be sufficiently icy.  
Evolution of main-belt objects onto JFC-like orbits may very well have occurred in previous dynamical studies investigating NEO origins, but simply not been noticed due to the evolution of $T_J$ values not being explicitly considered in those studies.  Therefore, while new large-scale and longer-term dynamical integrations to follow up the work presented here would be highly desirable, one simple next step could be to reexamine integration data sets generated by large-scale NEO population studies to identify asteroids that evolve onto low-$T_J$ orbits.

Meanwhile, theoretical modeling of the thermal evolution of objects evolving from the asteroid belt to JFC-like orbits would be useful for improving predictions about how many such objects could retain enough near-surface ice to support observable sublimation-driven activity. Thermal modeling aimed at placing constraints on the active lifetimes of these objects would also be useful for estimating the total number of objects from the Themis family on JFC-like orbits that could actually be active at any given time.

\section{Summary}\label{section:summary}

In this work, we present the results of numerical integrations of test particles representing members of the Themis asteroid family aimed at investigating dynamical pathways from potentially icy Themis family asteroids and the active Jupiter-family comet population.  We report the following key results:
\begin{enumerate}
    \item{Purely gravitational numerical integrations show that on the order of a few percent of current Themis family asteroids escape the family every 100 Myr and evolve onto orbits that are dynamically similar to those of Jupiter-family comets (JFCs).  If ice is widespread on Themis family objects as some independent studies have suggested and can be preserved in near-surface layers until those objects evolve onto JFC-like orbits, these dynamical results suggest that these objects have the potential to become active and thus mimic JFCs from the outer solar system despite originating in the main asteroid belt.}
    \item{The dynamical pathway by which most Themis family asteroids reach JFC-like orbits in our integrations is one in which objects near the 2:1 mean-motion resonance with Jupiter have their eccentricities and inclinations excited by the resonance, which then causes their orbits to approach and cross those of the major planets, particularly the terrestrial planets, Jupiter, and Saturn.  Many of these objects subsequently experience close encounters with these planets, which appear to help to drive their Tisserand parameter values with respect to Jupiter from asteroid-like values ($T_J>3$) down to JFC-like values ($2<T_J<3$).}
    \item{We estimate that, at any given time, there may be tens of objects from the Themis family on JFC-like orbits with the potential to mimic active JFCs from the outer solar system (compared to a total population of $\sim$600 currently known JFCs), although not all, or even any, may necessarily be observably active.  Calculations based on our purely gravitational integration results indicate that $\sim$20 objects with $D\geq1$~km from the Themis family may be on JFC-like orbits at any given time, but given uncertainties in the size of the relevant source population in the Themis family and the precise effects of factors like non-gravitational forces and collisional perturbations, we regard this simply as an order of magnitude estimate of the number of dynamically evolved Themis family objects with the potential to mimic active JFCS.
    %Meanwhile, the behavior of dynamical clones that are slightly offset in orbital element space from real Themis family asteroids suggests that small impulsive perturbations, such as impacts, could also help to increase the expected flux of Themis family asteroids onto JFC-like orbits.
    %\addedtext{Given these and other complicating factors, we consider our estimate of $\sim$20 objects from the Themis family on JFC-like orbits at any given time computed from our purely gravitational integrations alone to be meaningful to an order of magnitude at best, and thus conclude that there may be tens of such objects at any given time (compared to a total population of $\sim$600 currently known JFCs) that are potentially capable of exhibiting cometary activity.}
    }
    \item{Dynamically evolved Themis family asteroids that reach JFC-like orbits have orbital element distributions while on such orbits that are largely compatible with those of real JFCs, indicating that they could plausibly infiltrate the JFC population without being obvious dynamical outliers.  One distinguishing feature that could help to identify these objects is 
    %the strong clustering of their semimajor axes in the vicinity of the 2:1 mean-motion resonance with Jupiter (i.e., $3.15~{\rm au} < a < 3.40~{\rm au}$). 
    that they have semimajor axes within 0.125~au from the 2:1 mean-motion resonance with Jupiter (i.e., between 3.15~au and 3.40~au) for the vast majority ($\sim$90\%) of their time on such orbits, consistent with the strong role that this resonance plays in their dynamical evolution.
    That said, we find indications that longer integrations could change both the predicted abundance and orbital element distribution of dynamically evolved Themis family asteroids on JFC-like orbits.}
    %\item{\addedtext{We conclude that plausible dynamical pathways exist from the Themis asteroid family to JFC-like orbits, and so the possibility that some currently known active JFCs may be from the Themis family rather than from the outer solar system exists.  Moreover, other regions of the outer asteroid belt may also contribute to the active JFC population.  Larger-scale and longer-term dynamical integrations sampling more of the outer asteroid belt would be very useful for investigating these possibilities.}}
\end{enumerate}

Integration data are available upon request by contacting the corresponding author.

\acknowledgments
We thank an anonymous reviewer for helpful comments that improved this paper.  HHH, KW, and NS acknowledge support from NASA via the Solar System Workings program (Grant 80NSSC17K0723) as well as Solar System Exploration Research Virtual Institute (SSERVI) Cooperative Agreement grant NNH16ZDA001N.  BN acknowledges support by the Ministry of Education, Science and Technological Development of the Republic of Serbia, Project 176011.  The authors acknowledge the PSI Computing Center (PSICC) for providing HPC resources that contributed to the research results reported here.  %All figures have been plotted using {\tt matplotlib} \citep{hunter2007_matplotlib}.

\software{{\tt matplotlib} \citep{hunter2007_matplotlib},
    {\tt mercury} \citep{chambers1999_mercury},
    {\tt orbfit} and {\tt ORBIT9} \citep[][{\tt http://adams.dm.unipi.it/orbfit/}]{milani1988_orbit9}
          }

\clearpage

% BibTeX users please use one of
%\bibliographystyle{icarus}
%\bibliographystyle{elsarticle-harv}
%\bibliographystyle{abbrvnat}
\bibliographystyle{aasjournal}
\bibliography{themis-integrations}   % name your BibTeX data base

\begin{thebibliography}{}
\expandafter\ifx\csname natexlab\endcsname\relax\def\natexlab#1{#1}\fi
\providecommand{\url}[1]{\href{#1}{#1}}

\bibitem[{{A'Hearn} {et~al.}(1995){A'Hearn}, {Millis}, {Schleicher}, {Osip}, \&
  {Birch}}]{ahearn1995_ensemblecomets}
{A'Hearn}, M.~F., {Millis}, R.~C., {Schleicher}, D.~O., {Osip}, D.~J., \&
  {Birch}, P.~V. 1995, \icarus, 118, 223

\bibitem[{{Bottke} {et~al.}(2005){Bottke}, {Durda}, {Nesvorn{\'y}}, {Jedicke},
  {Morbidelli}, {Vokrouhlick{\'y}}, \& {Levison}}]{bottke2005_sizedistribution}
{Bottke}, W.~F., {Durda}, D.~D., {Nesvorn{\'y}}, D., {et~al.} 2005, \icarus,
  175, 111

\bibitem[{{Bottke} {et~al.}(2002){Bottke}, {Morbidelli}, {Jedicke}, {Petit},
  {Levison}, {Michel}, \& {Metcalfe}}]{bottke2002_neodistribution}
{Bottke}, W.~F., {Morbidelli}, A., {Jedicke}, R., {et~al.} 2002, \icarus, 156,
  399

\bibitem[{{Bottke} {et~al.}(2006){Bottke}, {Vokrouhlick{\'y}}, {Rubincam}, \&
  {Nesvorn{\'y}}}]{bottke2006_yarkovsky}
{Bottke}, Jr., W.~F., {Vokrouhlick{\'y}}, D., {Rubincam}, D.~P., \&
  {Nesvorn{\'y}}, D. 2006, Annual Review of Earth and Planetary Sciences, 34,
  157

\bibitem[{{Campins} {et~al.}(2010){Campins}, {Hargrove}, {Pinilla-Alonso},
  {Howell}, {Kelley}, {Licandro}, {Moth{\'e}-Diniz}, {Fern{\'a}ndez}, \&
  {Ziffer}}]{campins2010_themis}
{Campins}, H., {Hargrove}, K., {Pinilla-Alonso}, N., {et~al.} 2010, \nat, 464,
  1320

\bibitem[{{Chambers}(1999)}]{chambers1999_mercury}
{Chambers}, J.~E. 1999, \mnras, 304, 793

\bibitem[{{Cochran} {et~al.}(2012){Cochran}, {Barker}, \&
  {Gray}}]{cochran2012_cometspectroscopy}
{Cochran}, A.~L., {Barker}, E.~S., \& {Gray}, C.~L. 2012, \icarus, 218, 144

\bibitem[{{Di Sisto} {et~al.}(2005){Di Sisto}, {Brunini}, {Dirani}, \&
  {Orellana}}]{disisto2005_hildajfcs}
{Di Sisto}, R.~P., {Brunini}, A., {Dirani}, L.~D., \& {Orellana}, R.~B. 2005,
  \icarus, 174, 81

\bibitem[{{Di Sisto} {et~al.}(2019){Di Sisto}, {Ramos}, \&
  {Gallardo}}]{disisto2019_trojanevolution}
{Di Sisto}, R.~P., {Ramos}, X.~S., \& {Gallardo}, T. 2019, \icarus, 319, 828

\bibitem[{{Fern{\'a}ndez} \& {Sosa}(2015)}]{fernandez2015_jfcinterlopers}
{Fern{\'a}ndez}, J.~A., \& {Sosa}, A. 2015, \planss, 118, 14

\bibitem[{{Fern{\'a}ndez} {et~al.}(2000){Fern{\'a}ndez}, {Lisse}, {Ulrich
  K{\"a}ufl}, {Peschke}, {Weaver}, {A'Hearn}, {Lamy}, {Livengood}, \&
  {Kostiuk}}]{fernandez2000_encke}
{Fern{\'a}ndez}, Y.~R., {Lisse}, C.~M., {Ulrich K{\"a}ufl}, H., {et~al.} 2000,
  \icarus, 147, 145

\bibitem[{{Fern{\'a}ndez} {et~al.}(2013){Fern{\'a}ndez}, {Kelley}, {Lamy},
  {Toth}, {Groussin}, {Lisse}, {A'Hearn}, {Bauer}, {Campins}, {Fitzsimmons},
  {Licandro}, {Lowry}, {Meech}, {Pittichov{\'a}}, {Reach}, {Snodgrass}, \&
  {Weaver}}]{fernandez2013_jfcnuclei}
{Fern{\'a}ndez}, Y.~R., {Kelley}, M.~S., {Lamy}, P.~L., {et~al.} 2013, \icarus,
  226, 1138

\bibitem[{{Fink}(2009)}]{fink2009_cometsurvey}
{Fink}, U. 2009, \icarus, 201, 311

\bibitem[{{Granvik} {et~al.}(2017){Granvik}, {Morbidelli}, {Vokrouhlick{\'y}},
  {Bottke}, {Nesvorn{\'y}}, \& {Jedicke}}]{granvik2017_asteroidescape}
{Granvik}, M., {Morbidelli}, A., {Vokrouhlick{\'y}}, D., {et~al.} 2017, \aap,
  598, A52

\bibitem[{{Granvik} {et~al.}(2018){Granvik}, {Morbidelli}, {Jedicke}, {Bolin},
  {Bottke}, {Beshore}, {Vokrouhlick{\'y}}, {Nesvorn{\'y}}, \&
  {Michel}}]{granvik2018_neodistribution}
{Granvik}, M., {Morbidelli}, A., {Jedicke}, R., {et~al.} 2018, \icarus, 312,
  181

\bibitem[{{Greenstreet} {et~al.}(2012){Greenstreet}, {Ngo}, \&
  {Gladman}}]{greenstreet2012_neodistribution}
{Greenstreet}, S., {Ngo}, H., \& {Gladman}, B. 2012, \icarus, 217, 355

\bibitem[{{Haghighipour}(2009)}]{haghighipour2009_mbcorigins}
{Haghighipour}, N. 2009, Meteoritics and Planetary Science, 44, 1863

\bibitem[{{Hargrove} {et~al.}(2015){Hargrove}, {Emery}, {Campins}, \&
  {Kelley}}]{hargrove2015_antiope}
{Hargrove}, K.~D., {Emery}, J.~P., {Campins}, H., \& {Kelley}, M.~S.~P. 2015,
  \icarus, 254, 150

\bibitem[{{Horner} \& {Lykawka}(2010)}]{horner2010_trojanjfcs}
{Horner}, J., \& {Lykawka}, P.~S. 2010, International Journal of Astrobiology,
  9, 227

\bibitem[{{Hsieh} \& {Haghighipour}(2016)}]{hsieh2016_tisserand}
{Hsieh}, H.~H., \& {Haghighipour}, N. 2016, \icarus, 277, 19

\bibitem[{{Hsieh} \& {Jewitt}(2006)}]{hsieh2006_mbcs}
{Hsieh}, H.~H., \& {Jewitt}, D. 2006, Science, 312, 561

\bibitem[{{Hsieh} {et~al.}(2018){Hsieh}, {Novakovi{\'c}}, {Kim}, \&
  {Brasser}}]{hsieh2018_activeastfamilies}
{Hsieh}, H.~H., {Novakovi{\'c}}, B., {Kim}, Y., \& {Brasser}, R. 2018, \aj,
  155, 96

\bibitem[{{Hsieh} {et~al.}(2012){Hsieh}, {Yang}, {Haghighipour}, {Kaluna},
  {Fitzsimmons}, {Denneau}, {Novakovi{\'c}}, {Jedicke}, {Wainscoat},
  {Armstrong}, {Duddy}, {Lowry}, {Trujillo}, {Micheli}, {Keane}, {Urban},
  {Riesen}, {Meech}, {Abe}, {Cheng}, {Chen}, {Granvik}, {Grav}, {Ip},
  {Kinoshita}, {Kleyna}, {Lacerda}, {Lister}, {Milani}, {Tholen}, {Vere{\v s}},
  {Lisse}, {Kelley}, {Fern{\'a}ndez}, {Bhatt}, {Sahu}, {Kaiser}, {Chambers},
  {Hodapp}, {Magnier}, {Price}, \& {Tonry}}]{hsieh2012_288p}
{Hsieh}, H.~H., {Yang}, B., {Haghighipour}, N., {et~al.} 2012, \apjl, 748, L15

\bibitem[{Hunter(2007)}]{hunter2007_matplotlib}
Hunter, J.~D. 2007, Computing in Science \& Engineering, 9, 90

\bibitem[{{Jewitt}(1991)}]{jewitt1991_cometphotometry_cometsposthalley}
{Jewitt}, D. 1991, Astrophysics and Space Science Library, Vol. 167, {Cometary
  Photometry} (Springer Netherlands), 19--65

\bibitem[{{Kiselev} {et~al.}(2002){Kiselev}, {Jockers}, \&
  {Rosenbush}}]{kiselev2002_cometpolarimetry}
{Kiselev}, N., {Jockers}, K., \& {Rosenbush}, V. 2002, Earth Moon and Planets,
  90, 167

\bibitem[{{Kres{\'a}k}(1972)}]{kresak1972_tisserand}
{Kres{\'a}k}, L. 1972, in IAU Symposium, Vol.~45, IAU Symposium: The Motion,
  Evolution of Orbits, and Origin of Comets, 503

\bibitem[{{Lamy} {et~al.}(2004){Lamy}, {Toth}, {Fernandez}, \&
  {Weaver}}]{lamy2004_cometnuclei_comets2}
{Lamy}, P.~L., {Toth}, I., {Fernandez}, Y.~R., \& {Weaver}, H.~A. 2004, {The
  sizes, shapes, albedos, and colors of cometary nuclei} (University of Arizona
  Press, Tucson), 223

\bibitem[{{Levison} {et~al.}(1997){Levison}, {Shoemaker}, \&
  {Shoemaker}}]{levison1997_trojanevolution}
{Levison}, H.~F., {Shoemaker}, E.~M., \& {Shoemaker}, C.~S. 1997, \nat, 385, 42

\bibitem[{{Licandro} {et~al.}(2003){Licandro}, {Campins}, {Hergenrother}, \&
  {Lara}}]{licandro2003_124p}
{Licandro}, J., {Campins}, H., {Hergenrother}, C., \& {Lara}, L.~M. 2003, \aap,
  398, L45

\bibitem[{{Masiero} {et~al.}(2015){Masiero}, {DeMeo}, {Kasuga}, \&
  {Parker}}]{masiero2015_astfamilies_ast4}
{Masiero}, J.~R., {DeMeo}, F.~E., {Kasuga}, T., \& {Parker}, A.~H. 2015,
  Asteroids IV (Tucson, University of Arizona Press), 323--340

\bibitem[{{Michtchenko} {et~al.}(2016){Michtchenko}, {Lazzaro}, \&
  {Carvano}}]{michtchenko2016_mbadynamics}
{Michtchenko}, T.~A., {Lazzaro}, D., \& {Carvano}, J.~M. 2016, \aap, 588, A11

\bibitem[{{Milani} \& {Nobili}(1988)}]{milani1988_orbit9}
{Milani}, A., \& {Nobili}, A.~M. 1988, Celestial Mechanics, 43, 1

\bibitem[{{Morbidelli} {et~al.}(1995){Morbidelli}, {Zappala}, {Moons},
  {Cellino}, \& {Gonczi}}]{morbidelli1995_familyresonances}
{Morbidelli}, A., {Zappala}, V., {Moons}, M., {Cellino}, A., \& {Gonczi}, R.
  1995, \icarus, 118, 132

\bibitem[{{Murray} \& {Dermott}(2000)}]{murray2000_solarsystemdynamics}
{Murray}, C.~D., \& {Dermott}, S.~F. 2000, Solar System Dynamics (Cambridge
  University Press).
\newblock \url{https://www.xarg.org/ref/a/0521575974/}

\bibitem[{{Nesvorny}(2015)}]{nesvorny2015_pdsastfam}
{Nesvorny}, D. 2015, NASA Planetary Data System,
  EAR-A-VARGBDET-5-NESVORNYFAM-V3.0

\bibitem[{{Nesvorn{\'y}} {et~al.}(2017){Nesvorn{\'y}}, {Vokrouhlick{\'y}},
  {Dones}, {Levison}, {Kaib}, \& {Morbidelli}}]{nesvorny2017_spcorigin}
{Nesvorn{\'y}}, D., {Vokrouhlick{\'y}}, D., {Dones}, L., {et~al.} 2017, \apj,
  845, 27

\bibitem[{{Rivkin} \& {Emery}(2010)}]{rivkin2010_themis}
{Rivkin}, A.~S., \& {Emery}, J.~P. 2010, \nat, 464, 1322

\bibitem[{{Schleicher} {et~al.}(1993){Schleicher}, {Bus}, \&
  {Osip}}]{schleicher1993_43pwolfharrington}
{Schleicher}, D.~G., {Bus}, S.~J., \& {Osip}, D.~J. 1993, \icarus, 104, 157

\bibitem[{{Sch{\"o}rghofer}(2008)}]{schorghofer2008_mbaice}
{Sch{\"o}rghofer}, N. 2008, \apj, 682, 697

\bibitem[{{Sch{\"o}rghofer}(2016)}]{schorghofer2016_asteroidice}
---. 2016, \icarus, 276, 88

\bibitem[{Sch{\"o}rghofer \& Hsieh(2018)}]{schorghofer2018_asteroidiceloss}
Sch{\"o}rghofer, N., \& Hsieh, H.~H. 2018, Journal of Geophysical Research:
  Planets, 123, 2322

\bibitem[{{Sekanina}(1997)}]{sekanina1997_splitcomets}
{Sekanina}, Z. 1997, \aap, 318, L5

\bibitem[{{Snodgrass} {et~al.}(2006){Snodgrass}, {Lowry}, \&
  {Fitzsimmons}}]{snodgrass2006_cometphotometry}
{Snodgrass}, C., {Lowry}, S.~C., \& {Fitzsimmons}, A. 2006, \mnras, 373, 1590

\bibitem[{{Spoto} {et~al.}(2015){Spoto}, {Milani}, \& {Kne{\v
  z}evi{\'c}}}]{spoto2015_astfamages}
{Spoto}, F., {Milani}, A., \& {Kne{\v z}evi{\'c}}, Z. 2015, \icarus, 257, 275

\bibitem[{{Tancredi}(2014)}]{tancredi2014_asteroidcometclassification}
{Tancredi}, G. 2014, \icarus, 234, 66

\bibitem[{{Vokrouhlick{\'y}} {et~al.}(2006){Vokrouhlick{\'y}}, {Bro{\v z}},
  {Bottke}, {Nesvorn{\'y}}, \&
  {Morbidelli}}]{vokrouhlicky2006_yarkovskyfamilies}
{Vokrouhlick{\'y}}, D., {Bro{\v z}}, M., {Bottke}, W.~F., {Nesvorn{\'y}}, D.,
  \& {Morbidelli}, A. 2006, \icarus, 182, 118

\bibitem[{{Volk} \& {Malhotra}(2008)}]{volk2008_jfcsource}
{Volk}, K., \& {Malhotra}, R. 2008, \apj, 687, 714

\bibitem[{{Walsh} {et~al.}(2013){Walsh}, {Delb{\'o}}, {Bottke},
  {Vokrouhlick{\'y}}, \& {Lauretta}}]{walsh2013_eulaliapolana}
{Walsh}, K.~J., {Delb{\'o}}, M., {Bottke}, W.~F., {Vokrouhlick{\'y}}, D., \&
  {Lauretta}, D.~S. 2013, \icarus, 225, 283

\bibitem[{{Wang} \& {Malhotra}(2017)}]{wang2017_mmrs}
{Wang}, X., \& {Malhotra}, R. 2017, \aj, 154, 20

\end{thebibliography}

\end{document}